\documentstyle[12pt]{article}

\catcode`@=11
\def\un#1{\relax\ifmmode\@@underline#1\else
        $\@@underline{\hbox{#1}}$\relax\fi}
\catcode`@=12

\def\a{\alpha}
\def\b{\beta}
\def\c{\chi}
\def\d{\delta}

\def\f{\phi}
\def\g{\gamma}
\def\h{\eta}

\def\j{\psi}

\def\l{\lambda}
\def\m{\mu}
\def\n{\nu}

\def\p{\pi}

\def\r{\rho}
\def\s{\sigma}

\def\x{\xi}

\def\O{\Omega}

\def\S{\Sigma}

\def\ve{\varepsilon}
\def\vf{\varphi}

\def\bo{{\raise-.5ex\hbox{\large$\Box$}}}               
\def\pa{\partial}                                       
\def\TH{{\raise.2ex\hbox{$\displaystyle \bigodot$}\mskip-4.7mu \llap H \;}}
\def\face{{\raise.2ex\hbox{$\displaystyle \bigodot$}\mskip-2.2mu \llap {$\ddot
        \smile$}}}                                      
\def\dg{\sp\dagger}                                     

\def\sp#1{{}^{#1}}                              
\def\VEV#1{\left\langle #1\right\rangle}        
\def\abs#1{\left| #1\right|}                    
\def\leftrightarrowfill{$\mathsurround=0pt \mathord\leftarrow \mkern-6mu
        \cleaders\hbox{$\mkern-2mu \mathord- \mkern-2mu$}\hfill
        \mkern-6mu \mathord\rightarrow$}
\def\dvec#1{\vbox{\ialign{##\crcr
        \leftrightarrowfill\crcr\noalign{\kern-1pt\nointerlineskip}
        $\hfil\displaystyle{#1}\hfil$\crcr}}}           

\def\frac#1#2{{\textstyle{#1\over\vphantom2\smash{\raise.20ex
        \hbox{$\scriptstyle{#2}$}}}}}                   
\def\sfrac#1#2{{\vphantom1\smash{\lower.5ex\hbox{\small$#1$}}\over
        \vphantom1\smash{\raise.4ex\hbox{\small$#2$}}}} 
\def\bfrac#1#2{{\vphantom1\smash{\lower.5ex\hbox{$#1$}}\over
        \vphantom1\smash{\raise.3ex\hbox{$#2$}}}}       
\def\afrac#1#2{{\vphantom1\smash{\lower.5ex\hbox{$#1$}}\over#2}}    

\def\[{\lfloor{\hskip 0.35pt}\!\!\!\lceil}
\def\]{\rfloor{\hskip 0.35pt}\!\!\!\rceil}

\def\ha{{\fracmm12}}

\def\un{\underline}
\def\fracmm#1#2{{{#1}\over{#2}}}

\def\low#1{{\raise -3pt\hbox{${\hskip 0.75pt}\!_{#1}$}}}

\newskip\humongous \humongous=0pt plus 1000pt minus 1000pt

\newif\ifdtup

\def\ref#1{$\sp{#1)}$}

\parskip=\medskipamount                 
\thicklines                         

\begin{document}




\thispagestyle{empty}               

\def\border{                                            
        \setlength{\unitlength}{1mm}
        \newcount\xco
        \newcount\yco
        \xco=-24
        \yco=12
        \begin{picture}(140,0)
        \put(-20,11){\tiny Institut f\"ur Theoretische Physik Universit\"at
Hannover~~ Institut f\"ur Theoretische Physik Universit\"at Hannover~~
Institut f\"ur Theoretische Physik Hannover}
        \put(-20,-241.5){\tiny Institut f\"ur Theoretische Physik Universit\"at
Hannover~~ Institut f\"ur Theoretische Physik Universit\"at Hannover~~
Institut f\"ur Theoretische Physik Hannover}
        \end{picture}
        \par\vskip-8mm}

\def\headpic{                                           
        \indent
        \setlength{\unitlength}{.8mm}
        \thinlines
        \par
        \begin{picture}(29,16)
        \put(75,16){\line(1,0){4}}
        \put(80,16){\line(1,0){4}}
        \put(85,16){\line(1,0){4}}
        \put(92,16){\line(1,0){4}}

        \put(85,0){\line(1,0){4}}
        \put(89,8){\line(1,0){3}}
        \put(92,0){\line(1,0){4}}

        \put(85,0){\line(0,1){16}}
        \put(96,0){\line(0,1){16}}
        \put(79,0){\line(0,1){16}}
        \put(80,0){\line(0,1){16}}
        \put(89,0){\line(0,1){16}}
        \put(92,0){\line(0,1){16}}
        \put(79,16){\oval(8,32)[bl]}
        \put(80,16){\oval(8,32)[br]}

        \end{picture}
        \par\vskip-6.5mm
        \thicklines}

\border\headpic {\hbox to\hsize{
\vbox{\noindent ITP--UH--25/95 \hfill October 1995 \\
hep-th/9510005 \hfill }}}

\noindent
\vskip1.3cm
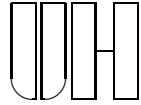
\begin{center}

{\Large\bf N=2 String Loops and Spectral Flow}~\footnote{Talk given at the
International Workshop on Strings, Gravity and Related Topics,
\newline ${~~~~~}$ June 29--30, 1995, ICTP, Trieste, Italy;
to appear in the Proceedings}

\vglue.3in
Sergei V. Ketov \footnote{Supported in part
by the `Deutsche Forschungsgemeinschaft'}\\
\vglue.1in
{\it Institut f\"ur Theoretische Physik, Universit\"at Hannover}\\
{\it Appelstra\ss{}e 2, 30167 Hannover, Germany}\\
{\sl ketov@itp.uni-hannover.de}
\end{center}

\vglue.5in
\begin{center}
{\Large\bf Abstract}
\end{center}
Based on our previous studies of the BRST cohomology of the critical
$N=2$ strings, we construct the loop measure and make explicit the role of the
spectral flow at arbitrary genus and Chern class, in a holomorphic field basis.
The spectral flow operator attributes to the existence of the hidden `small'
$N=4$ superconformal symmetry which is non-linearly realized. We also discuss
the symmetry properties of $N=2$ string amplitudes on locally-flat backgrounds.

\newpage

\section{Introduction}

The $N=2$ strings, i.e. strings with two world-sheet supersymmetries, have a
long history.~\cite{book} The $N=2$
critical strings are known to live in four dimensions, with the unphysical
signature $(2,2)$. The (smooth) target space can be e.g., ${\bf C}^{(1,1)}$,
${\bf T}^2\otimes {\bf R}^2$, ${\bf T}^{(2,2)}$, or any K\"ahler and
Ricci-flat Riemannian manifold with the signature $(2,2)$. The NSR-type
gauge-invariant world-sheet action is given by coupling the $N=2$ world-sheet
supergravity multiplet $\left(e^a_{\a},\c^{\pm}_{\a},A_{\a}\right)$ to two
complex $N=2$ scalar matter multiplets $(X^{\pm},\j^{\pm})^{\m}$, where
$\a=1,2,$ and $\m=0,1$.
The $N=2$ superconformal BRST gauge fixing produces (anti-commuting) conformal
ghosts $(b,c)_2$, complex (commuting) $N=2$ superconformal ghosts
$(\b^{\pm},\g^{\pm})_{3/2}$, and (anti-commuting) real $U(1)$ ghosts
$(\tilde{b},\tilde{c})_1$, where the spins appear as subscripts, and the
$U(1)$ charges $\pm$ as superscripts. The non-anomalous $N=2$ superconformal
algebra is generated by the total (with ghosts) stress tensor $T_{\rm tot}$,
the supercurrents $G^{\pm}_{\rm tot}$ and the $U(1)$ current
$J_{\rm tot}$.~\cite{klp}

Contrary to the $N=1$ string, chiral bosonization of the $U(1)$-charged NSR
fermionic fields $\j$ and ghosts $(\b,\g)$ depends on the field basis. In the
`real' basis,~\cite{pope} one bosonizes real and imaginary parts of that
fields, whereas it is their holomorphic and antiholomorphic
combinations that are taken in the `holomorphic' basis. We are going to work
in the holomorphic basis, since it has the advantage of diagonalizing the
local $U(1)$ symmetry. In the bosonized form, the  $\j^{\m,\pm}$ and
$(\b^{\pm},\g^{\pm})$ are replaced by two pairs of bosons ($\f^{\pm},
\vf^{\pm})$  and two auxiliary fermion ghost systems $(\h^{\pm},\x^{\pm})$,
which altogether form an extended Fock space of states containing
${\bf Z}\times{\bf Z}$ copies of the original space of fermionic states.

The natural global continuous (`Lorentz') symmetry of a flat
$(2+2)$-dimensional target space is $SO(2,2)\cong SU(1,1)\otimes SU(1,1)$,
but the NSR-type $N=2$
string action has only a part of it, namely $U(1,1)\cong U(1)\otimes SU(1,1)$.
\cite{klp} The spectrum of the critical $N=2$ string is given by the
cohomology of its BRST charge $Q_{\rm BRST}$,~\cite{bkl} and
it has only a finite number of massless physical states, all having vanishing
conformal dimension and $U(1)$ charge. Further grading of the cohomology is
effected by the total ghost number $u\in {\bf Z}$, and two picture numbers
$\p^{\pm}\in \ha {\bf Z}$, with $\p^+ + \p^-\in {\bf Z}$. The BRST cohomology
for generic momenta consists of four classes of states for each pair
 $(\p^+,\p^-)$, labelled by $v\equiv u-\p^+ -\p^-\in \{1,2,2',3\]$ and
created by vertex operators $V$ of the types $cW$, $\tilde{c}cW$, $c\pa c W$
and $\tilde{c}c\pa c W$, with ghost-independent $W$.~\cite{bkl}

Physical states are given by the equivalence classes of the BRST cohomology
classes under the following four equivalence relations. First, $cW$ and
$c\pa cW$ are to be identified just as in the bosonic string theory. Second,
$\tilde{c}$-type vertices get converted to others by applying the $U(1)$ ghost
number-changing operator $Z^0$. Third, two picture-changing operators
$Z^{\pm}$ raise the picture numbers of vertices by unit amounts. And, fourth,
NS- and R-type states are connected by the spectral flow~\cite{ov}
$SFO^{\pm}$, which move $(\p^+,\p^-)\to (\p^+\pm\ha,\p^-\mp\ha)
$. Explicitly, these maps are given by \cite{kl}
$$Z^0=\oint \tilde{b}\,\d\left(\oint J_{\rm tot}\right)~,\quad
Z^{\pm}(z)=\d(\b^{\pm})G^{\pm}_{\rm tot}~,\quad SFO^{\pm}(z)= \exp \left(
 \pm\ha \int^z J_{\rm tot}\right)~,\eqno(1)$$
they commute with $Q_{\rm BRST}$ but are non-trivial. In this fashion, each
physical state has a representative $\tilde{c}cW_{\rm can}$ at $v=2$, in the
canonical picture  $(\p^+,\p^-)=(-1,-1)$. Vertex operators with other ghost
and picture numbers are however needed for an actual computation of string
amplitudes, in order to meet certain selection rule requirements.~\cite{book}
As a net result,~\cite{bkl} only a single scalar excitation \cite{ov} survives
in an interacting theory, whereas the twisted (NS-R and R-NS) physical states,
which would-be the target space `fermions', decouple.~\footnote{There are also
{\it discrete} physical states at {\it vanishing} momenta. They are to be
accounted, if one wants to get correct factorization properties of $N=2$
string amplitudes at vanishing momenta.} In particular, there cannot be an
interacting NSR-type critical $N=2$ string model with a `space-time'
supersymmetry ({\it cf.} \cite{siegel}).

\section{$N=2$~ String Loops}

To compute any $n$-point amplitude, one needs to sum over all genera
$h\in {\bf Z}^+$ and $U(1)$ instanton numbers (Chern class) $c\in {\bf Z}$ of
the Euclidean world-sheet (punctured Riemann surface) $\S_{h,n}$, where
$$\c=\fracmm{1}{2\p}\int_{\S}R=2-2h~,\qquad c=\fracmm{1}{2\p}\int_{\S}F~,
\quad F=dA~.\eqno(2)$$
To compute the contribution for fixed $h$ and $c$, one must integrate
out $2h-2\pm c+n$ complex fermionic supermoduli of $U(1)$ charge $\pm 1$,
respectively, and $h-1+n$ complex $U(1)$ moduli, to obtain an integration
measure for the remaining $3h-3+n$ complex metric moduli.~\footnote{The cases
of the sphere and torus require some modifications, according to the index
theorems.} Among the complex metric moduli, $3h-3$ ones are associated with
holomorphic quadratic differentials which are dual to closed non-intersecting
geodesics on $\S$, and $n$ additional ones are the positions $(z_l)$ of vertex
operators (punctures). The supermoduli count solutions to the equation
$\hat{D}^{\pm}\c^{\pm}_z\equiv\left(\bar{\pa}\mp iA_{\bar{z}}\right)\c^{\pm}_z
=0$, whose number is dictated by the Riemann-Roch theorem:
$$ {\rm ind}\,\hat{D}^{\pm}\equiv {\rm dim}\,{\rm ker}\hat{D}^{\pm}
- {\rm dim}\,{\rm ker}\hat{D}^{\pm}{}^{\dg}=2(h-1) \pm c +n~.\eqno(3)$$
For $h>1$, the contributions to a positive index generically come from the
first term. When the index becomes negative, $\det\hat{D}^{\pm}{}^{\dg}$
develops zero modes, which implies the vanishing of the corresponding
correlation function. As a result, the amplitudes for $\abs{c}>2(h-1)+n$ all
vanish. The $U(1)$ moduli space can be seen as the moduli space of flat
connections on $\S_{h,n}$, and it is a product of two factors. One factor is
the Jacobian variety of flat $U(1)$ connections on $\S_{h,0}$, dual to the
homology,
$$J(\S_{h,0})=\fracmm{{\bf C}^h}{{\bf Z}^h+\O{\bf Z}^h}~,\eqno(4)$$
where $\O$ is the period matrix of $\S$. It is parametrized by the real twists
$\oint_{a_i} A$ and $\oint_{b_i} A$ around the homology cycles $a_i$ and $b_i$.
 The other factor is the torus ${\bf R}^{2n-2}/{\bf Z}^{2n-2}$, encoding the
$2n-2$ independent twists $\oint_{c_l} A$ and $\oint_{c_l}*A$ around the
punctures $z_l$. The associated homology `cycles' in the dual space are just
$n-1$ independent curves connecting the punctures with a reference point $z_0$.

 To get the proper gauge fixing (Faddeev-Popov determinant), one should
get rid of ghost zero modes by inserting anti-ghosts for each moduli direction,
which come paired with Beltrami differentials, similarly to the $N=1$
string.~\cite{vv} Since the total action is linear in the gauge fields and,
hence, in the supermoduli and $U(1)$ moduli, one gets the additional insertions
 of delta-functions of the corresponding total currents. Putting all
together, and using the identity $Z^{\pm}=\{Q_{\rm BRST},\x^{\pm}\}
=\d(\b^{\pm})G^{\pm}_{\rm tot}$, we get a product of the BRST-invariant $N=2$
 picture-changing operators, as expected ({\it cf.} \cite{bv}),
$$\VEV{ \abs{ \left(\oint b\right)^{3h-3+n}(Z^+)^{2h-2+c+n}(Z^-)^{2h-2-c+n}
(Z^0)^{n-1}}^2\prod_{i=1}^h\left[ Z^0(a_i)Z^0(b_i)\right]V^{\rm can}_1\cdots
V^{\rm can}_n}~.\eqno(5)$$
The picture-changing operators $Z^{\pm}$ and the $Z^0$ can be used to convert
vertex operators to other pictures and/or ghost numbers. In particular, the
$U(1)$ number-changing operators $Z^0$ in eq.~(5) enforce a projection onto
charge-neutral excitations propagating across handles, which guarantees a
factorization of the amplitude on neutral states only.~\footnote{The
instanton-number-changing operators $Z^0$ have no analogue in the $N=0$ and
$N=1$ strings.}

\section{Spectral Flow}

Invariance of correlation functions under the spectral flow follows from the
fact that a change in monodromies for the world-sheet fermions is equivalent to
a shift in the integration over $U(1)$ moduli.~\cite{ov} However, this changes
the loop measure by a factor $e^{-\langle\d A,J_{\rm tot}\rangle}$. Since
$\d A$ is harmonic away from the punctures $z_l$, and $d*J_{\rm tot}=0$, we
have
$$\langle \d A,J_{\rm tot} \rangle =
\sum^h_{i=1}\left( \oint_{a_i}\d A \oint_{b_i}*J_{\rm tot} -
\oint_{b_i}\d A \oint_{a_i}*J_{\rm tot}\right) +\sum^n_{l=1}\oint_{c_l}\d A
\int^{z_l}_{z_0}*J_{\rm tot}~.\eqno(6)$$
which results in the twists $e^{2\p i\l\int_{\g}*J_{\rm tot}}$ around the
cycles $\g=(a_i,b_i,c_l)$. The harmonicity of $\d A$ implies
$$\sum_l{\rm Res}_{z_l}\d A=0~,\quad {\rm or,~equivalently,~}\quad
 \sum^n_{l=1}\l_l=0~.\eqno(7)$$
In physical terms, $\l_l$ is just the flux of the monopole field in Dirac
string through $z_l$. On the vertices, the spectral flow is realized by
$$V_l(z_l)\longrightarrow V_l^{(\l_l)}(z_l)\equiv SFO(\l_l,z_l)
\cdot V_l(z_l)~,\eqno(8)$$
where the `spectral flow operator' takes the form~\cite{kl}
$$ SFO(\l,z)=\exp\left(2\p i\l \int^z_{z_0}*J_{\rm tot}\right)
= e^{2\p i\l\left(\f^+ -\f^- -\vf^+ +\f^-+\tilde{b}c\right)}~.\eqno(9)$$
Eq.~(7) ensures that the spectral flow operator does not depend on the
reference point $z_0$, it is a local~\footnote{{\it After} chiral
bosonization, formulations in the real and holomorphic basises are non-locally
related. Therefore, the spectral flow operator is non-local in the real basis.}
  ~and BRST-invariant operator. When $\l=\pm\ha$, it maps NS into R sector,
whereas for $\l\in{\bf Z}$ it maps from the $c$-instanton sector to the
$(c+\l)$-one. Its effect on the correlation functions is
$$\VEV{V_1^{(\l_1)}\cdots V_n^{(\l_n)}}=\VEV{V_1\cdots V_n\prod_lSFO(\l_l)}
\stackrel{\sum \l_l=0}{=} \VEV{V_1\cdots V_n}~.\eqno(10)$$
equating all $n$-point amplitudes with the same values for $h$ and $c$.

\section{Amplitudes and Partition Function}

Adding to the generators of the underlying $N=2$ superconformal algebra the
spectral flow operators $SFO(1,z)\equiv J^{++}(z)$ and
$SFO(-1,z)\equiv J^{--}(z)$, and closing the algebra, one arrives at the
`small\/' linear $N=4$ superconformal algebra. The presence of $Z^0$
insertions in the
loop measure has an effect of the topological twist. After the twist, the
$(\b^-,\g^-)$ ghosts acquire spin two, which is the same as that of the
$(b,c)$ system,  whereas the $(\b^+,\g^+)$ ghosts get spin one, just as that
of the $(\tilde{b},\tilde{c})$ system. Since the respective ghosts have the
opposite statistics, all the $N=2$ ghosts {\it cancel} after the twist, at
least in the partition function.~\cite{bv} This paves the way for treating the
$N=2$ string measure as that of an $N=4$ topological field theory.~\footnote{
There are indications~\cite{ovl} \/that the latter is the {\it holomorphic}
Yang-Mills theory in the limit  $N_{\bf c}\to\infty$.}

As far as the partition function is concerned $(n=0)$, $N=4$ topological
calculations can be put under control for the target space
${\bf T}^2\otimes{\bf R}^2$, since the results can be recognized as
topological invariants.~\cite{ovl} For example, the 1-loop partition function
is proportional to~\cite{ovl}
$-\log\left(\sqrt{{\rm Im}\s{\rm Im}\r}\abs{\h(\s)}^2\abs{\h(\r)}^2
\right)$, which is obviously mirror-symmetric under the exchange of the
K\"ahler and complex moduli ($\s$ and $\r$, respectively) of ${\bf T}^2$.
Similarly, the 2-loop partition function at the {\it maximal} instanton
numbers $(2,2)$ is $\s$-holomorphic, and is given by the Eisenstein series of
degree $4$.

As far as the scattering amplitudes are concerned, all the $n$-point functions
beyond $n=3$ are expected to vanish.~\cite{bv} The non-trivial example is
provided by the case of $n=3$ and $h=c=0$. The corresponding correlator is
most easily computed in the $(-2,-2)$ total picture,~\cite{bkl}
$$\VEV{V(k_1,z_1)V(k_2,z_2)V(k_3,z_3)}^{\rm left}_{0,0}\propto
\left(k^+_2\cdot k_3^- -k^-_2\cdot k_3^+\right)~,\eqno(11a)$$
and it is $U(1,1)$ invariant.~\cite{ov} The case of $n=3$, $h=0$ and $c=1$ is
a bit more complicated,
$$\VEV{\tilde{c}cW_{(-1,-1)}(k_1,z_1)cW_{(0,-1)}(k_2,z_2)cW_{(0,-1)}(k_3,z_3)}
^{\rm left}_{0,1}\propto
\ve^{\m\n}k^{\m-}_2 k_3^{\n -}~,\eqno(11b)$$
and it is only $SO(1,1)$ invariant. Our final example is the 1-loop $3$-point
function at $c=0$, which reads ({\it cf.} \cite{it})
$$\VEV{\int W_{(0,0)}\int W_{(0,0)}\int W_{(0,0)}}_{(1,0)}\propto
\left(k^{[+}_2\cdot k_3^{-]}\right)^3~,\eqno(12)$$
and it is non-vanishing, and local in momenta.

\section{Acknowledgements}

This work was done in collaboration with O. Lechtenfeld. Fruitful discussions
with N. Berkovits, E. Gava, H. Ooguri, C. Pope and C. Vafa are acknowledged.

\end{document}